\documentclass[iop,apj]{emulateapj}
\usepackage{graphicx}
\usepackage{float}
\usepackage{natbib}
\graphicspath{ {~} }
\begin{document}

\title{A Search For Pulsations in the Optical Light Curve of the Nova ASASSN-17hx}
\author{ Lee J. Rosenthal\altaffilmark{1}, Ken Shen\altaffilmark{2}, Gregg Hallinan\altaffilmark{1}, Navtej Singh\altaffilmark{3}, Laura Chomiuk\altaffilmark{4}, Raffaella Margutti\altaffilmark{5}, Brian Metzger\altaffilmark{6}}

\altaffiltext{1}{California Institute of Technology, 1200 E California Blvd, Pasadena, CA 91125, USA}
\altaffiltext{2}{Astronomy Department, University of California Berkeley, Berkeley, CA 94720, USA}
\altaffiltext{3}{Jet Propulsion Laboratory, 4800 Oak Grove Dr, Pasadena, CA 91109, USA}
\altaffiltext{4}{Department of Physics and Astronomy, Michigan State University, East Lansing, MI 48824, USA}
\altaffiltext{5}{Department of Physics and Astronomy, Northwestern University, Evanston, IL 60208, USA}
\altaffiltext{6}{Department of Physics and Columbia Astrophysics Laboratory, Columbia University, New York, NY 10027, USA}
\email{lrosenth@caltech.edu}

\begin{abstract}

We present high-speed optical observations of the nova ASASSN-17hx, taken both immediately after its discovery and close to its first peak in brightness, to search for seconds -- minutes pulsations associated with the convective eddy turnover timescale within the nova envelope. We do not detect any periodic signal with greater than $3\sigma$ significance. Through injection and recovery, we rule out periodic signals of fractional amplitude $>7.08\times10^{-4}$ on timescales of 2 seconds and fractional amplitude $>1.06\times10^{-3}$ on timescales of 10 minutes. Additional observations of novae are planned to further constrain ongoing simulations of the launch and propagation of nova winds.

\end{abstract}

\section{Introduction}

The consensus model for a classical nova invokes a binary star system, with accretion from a main-sequence star or evolved giant onto a white dwarf (WD) due to Roche lobe overflow (e.g., \citealt{CV-Review-Short}). As hydrogen-rich material is transferred to the WD through an accretion disk, the temperature at the base of the accreted envelope rises until it reaches $\sim2\times10^7$ K, at which point the accreted fuel undergoes fusion via the CNO cycle.  A convective zone is born and grows until an optically-thick wind is launched \citep{KatoHachisu} and the luminosity approaches the Eddington limit, giving rise to the observed classical nova.

The launching of the optically-thick wind is primarily due to the presence of the iron opacity bump \citep{KatoHachisu}.  While the total luminosity throughout the envelope remains somewhat sub-Eddington with respect to Thompson opacity, the luminosity becomes locally super-Eddington once the envelope has expanded and cooled sufficiently such that the local opacity is enhanced due to the iron opacity bump.

To demonstrate this evolution, we model a nova outburst on a $1.2 \, M_\odot$ WD accreting solar composition material at $10^{-9} \, M_\odot \, {\rm yr^{-1}}$ using the stellar evolution code \texttt{MESA}\footnote{\url{http://mesa.sourceforge.net}; version 8118} \citep{paxt11,paxt13,paxt15a,paxt18a}.  The appropriate hydrodynamic and boundary condition flags are implemented to follow the launch and propagation of a steady-state wind.  Further simulation details will be provided in Shen et al.\ (2018, in preparation).

\begin{figure}
\begin{center}
\includegraphics[width = \columnwidth]{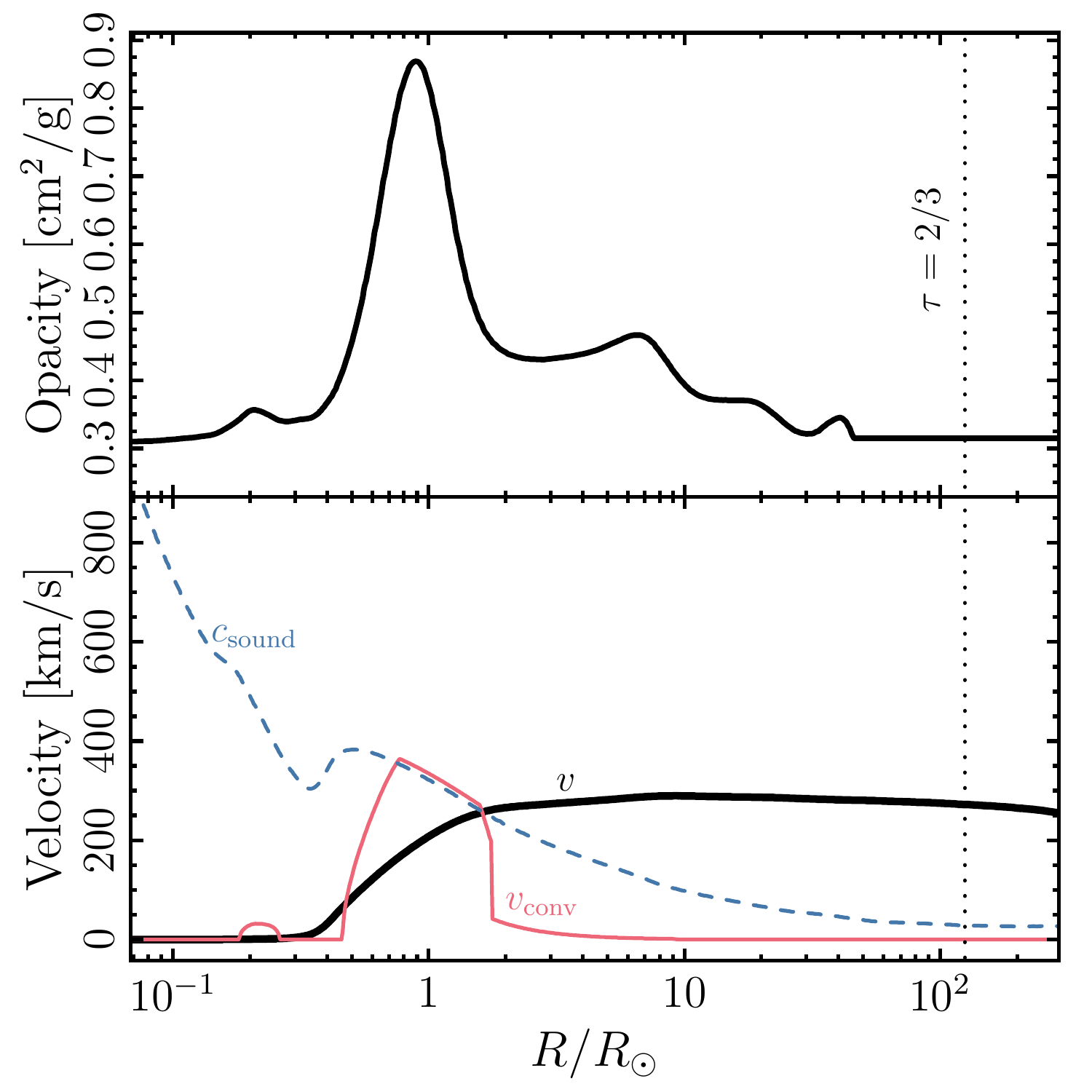}\\
\caption{Radial profiles of opacity (top panel), sound speed, $c_{\rm sound}$, convective velocity, $v_{\rm conv}$, and fluid velocity, $v$ (bottom panel) at the optical peak of our fiducial nova simulation.  The $\tau=2/3$ photosphere is indicated by a dotted line in both panels.  \label{fig:profiles}}
\end{center}
\end{figure}

Figure \ref{fig:profiles} shows radial profiles from this simulation of opacity (top panel), sound speed, convective velocity, and fluid velocity (bottom panel) at the nova's optical peak, $54 \, {\rm d}$ after the birth of the main convective zone, which is found between $0.1$ and $0.2 \, R_\odot$.  The $\tau=2/3$ photosphere is denoted by a dotted line.  A large iron opacity bump is clearly visible at $0.6 \, R_\odot$; it is here that a second convection zone develops in an attempt to carry the locally super-Eddington luminosity that results due to the large opacity increase. However, the convective velocities in the \texttt{MESA} simulation are constrained to be subsonic, and as a result, the only outlet for the super-Eddington luminosity is the launching of a wind.

While the second convective zone fails to transport the super-Eddington luminosity, it may still leave an observable imprint in the classical nova light curve.  The convective eddy turnover timescale at the peak of the convective velocity is $v_{\rm conv}/ H_P \sim 3 \, {\rm min}$, where $H_P$ is the pressure scale height.  These motions may produce significant temporal variability at this period, which could then be advected out towards the photosphere, where it may be detected with high-cadence photometric observations.  The precise frequency of these pulsations is uncertain; the peak velocity and its location vary as the nova evolves, and the convective velocity also varies within the convective zone.  Furthermore, it is not obvious how convective motions will be influenced by the accelerating envelope in the wind launching region.

Moreover we note here that recent evidence has supported a more complicated view of classical nova outbursts as being dominated by interaction with the donor (e.g., \citealt{chom14a}).  Mass loss through the binary's outer Lagrange point, similar to the mechanism proposed for outbursts from stellar mergers \citep{pejc16a}, may determine the initial evolution of the outburst; interaction between this slow, dense, equatorial outflow and a later, faster, more spherical ejecta may also explain recent detections of gamma-rays from classical novae \citep{abdo10a,acke14a,metz14a}.  This updated picture of nova outbursts, with alternative ejection mechanisms and luminosity sources, will muddy the classic scenario and may render the pulsations unobservable. Therefore, while a detection of a periodic signal would provide a new, powerful diagnostic of early time nova fireballs, a non-detection would remain ambiguous. Nonetheless, we have commenced the first systematic search for early time oscillations of a small sample of novae to place initial constraints on the presence of such periodic signals from timescales of $\sim 1$~second to 30 minutes, with initial focus on the Galactic nova ASASSN-17hx.

In Section 2 we introduce the Caltech High-Speed Multicolor Camera (CHIMERA), the instrument used to observe ASASSN-17hx, as well as instrument and observing conditions for our observations. In Section 3, we describe how we performed image reduction and differential photometry to extract the nova light curves, and statistical analysis to search for periodic signals in the light curves. In Section 4, we briefly outline the lack of a detection of periodicity, and infer constraints on the predicted amplitude of periodicity in nova brightness. In Section 5, we discuss the significance of our findings and potential future work to constrain second-timescale nova pulsations.

\section{Observations}

\subsection{ASASSN-17hx}

ASASSN-17hx is a Galactic nova that was detected on June 23rd UTC, 2017, by the All Sky Automated Survey for SuperNovae (ASASSN), a network of robotic telescopes designed to detect transients \citep{ASASSN-Init, ASASSNProject}.ASASSN-17hx was spectroscopically confirmed as a nova with observations from the Rozhen observatory in Bulgaria \citep{ASASSN-Spec}. In particular, it likely is a binary system with accretion from a main-sequence or giant companion onto a WD as its progenitor. While the initial spectroscopic confirmation in June identified ASASSN-17hx as a He/N nova, meaning that Helium and Nitrogen lines dominated the nova's optical spectrum, further spectroscopy done over the proceeding months \citep{Liverpool,Himalayas,ARAS}, revealed increasingly strong Fe II features. This fits the theory that most novae are hybrids of the He/N type and the Fe II type, with the Fe II features coming into prominence after peak brightness \citep{Classification}.

According to data compiled by the American Association of Variable Star Observers \citep{AAVSOdata}, the nova first peaked on July 27th, then declined and re-brightened from August to September. It peaked again on September 16th, and experienced a third, smaller peak in early October. Multiple peaks are common in classical novae, although their physical origin is still not fully understood \citep{BodeEvans}.

\subsection{CHIMERA}

The data analyzed in this paper have been acquired by the Caltech HIgh-speed Multicolor camERA (CHIMERA), a high-speed optical photometer operated in the prime focus of the Palomar 200-inch telescope \citep{CHIMERA}. CHIMERA is designed to monitor objects that fluctuate on timescales ranging from milliseconds to hours. Its field of view is 5 x 5 arcminutes, and when the entire CCD is read out, observations can have exposure times as low as 36 milliseconds. Windowing to smaller fields of view, or binning pixels, allows for even lower exposure times.

CHIMERA has a dichroic that separates incoming light into a red and a blue channel, each of which is serviced by its own Electron-Multiplying (EM) CCD. This enables simultaneous observations in two distinct optical bands. Each of these channels has a separate filter mechanism, one containing SDSS $u'$ and $g'$, and the other containing $r'$, $i'$, and $z'$. Each channel also has a clear throughput option, with properties defined by the dichroic and CCD throughputs. See Harding et al. 2016 for throughput details.

\subsection{Observations of ASASSN-17hx}

We observed ASASSN-17hx with CHIMERA on UTC June 25th, June 26th, July 22nd, and July 23rd, 2017. Observation details are included in Table 1. Figure 2 shows an example image with reticles centered on ASASSN-17hx and the stars we used for differential photometry, and Figure 3 shows a light curve of the nova, including our observing dates. The June nights were mostly photometric, with some periods of light cloud cover. There were heavier periods of cloud cover during our July nights.

\begin{table*}[t]
 \centering
  \begin{tabular}{ c || c | c | c | c | c }
  	Date & Red filter & Blue filter & Chip readout size & Exposure & Duration \\ \hline \hline
    6/25/17 & $r'$ & $g'$ & 1024x1024 & 1.1 s & 45 min \\
    6/26/17 & Clear & $g'$ & 512x512 & 0.55 s & 90 min \\
    7/22/17 & $r'$ & $g'$ & 512x512 & 0.55 s & 90 min \\
    7/23/17 & $r'$ & $g'$ & 256x256 & 1.65 s & 90 min
  \end{tabular}
  \caption{Observational parameters on each night.}
  \label{tab:1}
\end{table*}

\begin{figure}
\begin{center}
\includegraphics[width = \columnwidth]{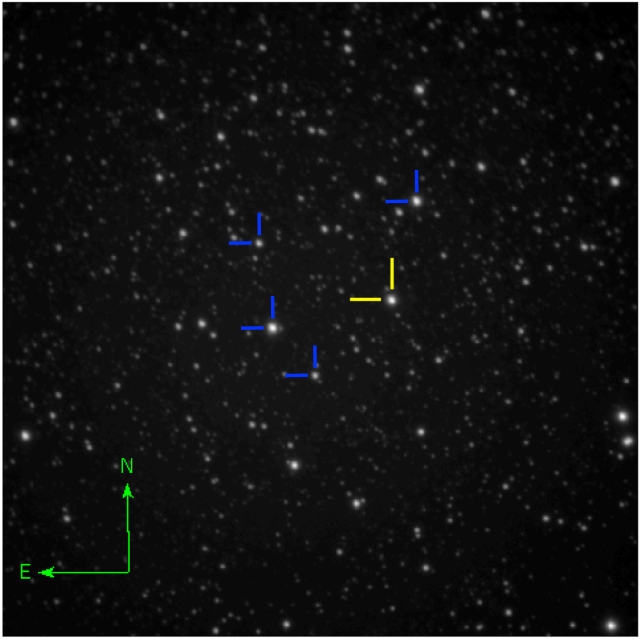}\\
\caption{Image of ASASSN-17hx taken on June 25th 2017. The yellow reticle designates the nova, while the smaller blue reticles designate our reference stars.}
\end{center}
\end{figure}

\begin{figure}
\begin{center}
\includegraphics[width = \columnwidth]{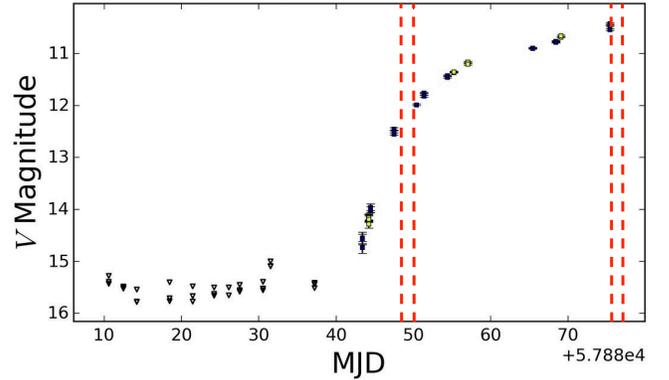}\\
\includegraphics[width = \columnwidth]{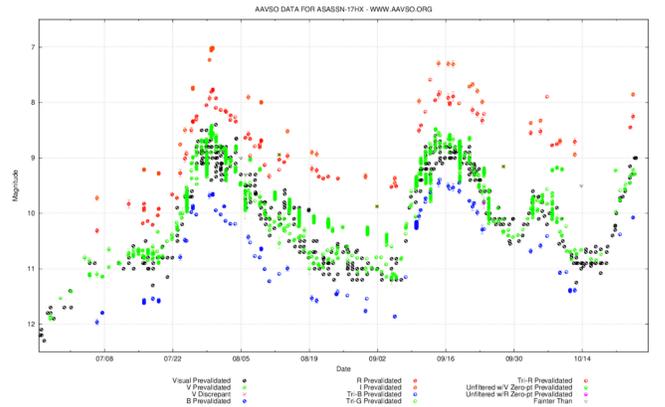}\\
\caption{Top panel: ASASSN light curve of ASASSN-17hx, courtesy of \cite{ASASSN-Init}. Vertical lines denote observing dates. Bottom panel: American Association of Variable Star Observers (AAVSO) light curve, spanning a longer period of time and showing multiple peaks \citep{AAVSOdata}.}
\end{center}
\end{figure}

\section{Methods}

\subsection{Calibration and differential photometry}

We performed standard de-biasing and flat-fielding of the data with the PyChimera\footnote{https://github.com/navtejsingh/pychimera} pipeline, utilizing Daophot and Pyraf \citep{Daophot}. We used 0.01 s bias frames, and twilight flats, to de-bias and flat-field our images. Because our exposure times do not exceed 1.1 seconds, dark current is negligible and ignored. We performed differential photometry on the nova relative to the brightest stars in the field of view in order to normalize flux against fluctuations caused by cloud cover, atmospheric conditions, changing airmass, and other effects separate from the intrinsic nova flux.  This field of view contains four stars sufficiently bright for use in differential photometry.

The baseline generated with our reference stars may be complicated by the cloud crossing time at Palomar, combined with the 0.55-1.1 second exposure times. The cloud crossing time is defined here as the limiting case of the average time it takes a cloud to cross CHIMERA's field of view at zenith. If the cloud crossing time is greater than the exposure time, the fluxes of stars on opposite sides of the CHIMERA field of view are impacted at different times, therefore differential photometry may not fully cancel out the effects of cloud cover. We estimate the cloud-crossing time as the time it takes for a cirrus cloud above Palomar to cross CHIMERA’s 5x5 arcminute field of view. Cirrus cloud speed falls within the range $\sim 45-71.5\frac{\textrm{m}}{\textrm{s}}$ \citep{ahrens2006meteorology}. We use the upper limit of this range in our calculation, to place a lower limit of 0.0874 seconds on the cloud crossing time.

If it takes cirrus clouds roughly 0.1 seconds to cross CHIMERA's field of view, then the time lag of their effects on the brightness of stars on opposite sides of the field of view might be significant in 0.5 second exposures. This appears to be the case in the first night of data, where a small dip in brightness due to cloud cover is still present in the normalized nova light curve. Aside from this fluctuation, our normalized data seem largely uncorrupted by cloud cover. As shown in the next section, we can confirm this finding by comparing the time-series power spectra of our reference star light curves to the power spectra of our nova light curves. If, on a given night, the reference stars share strong frequency peaks with that night's nova power spectrum, then these frequency peaks may be caused by cloud contamination. We plot an example of differential light curves in Figure 4.

\begin{figure}
\begin{center}
\includegraphics[width = 0.49\textwidth]
{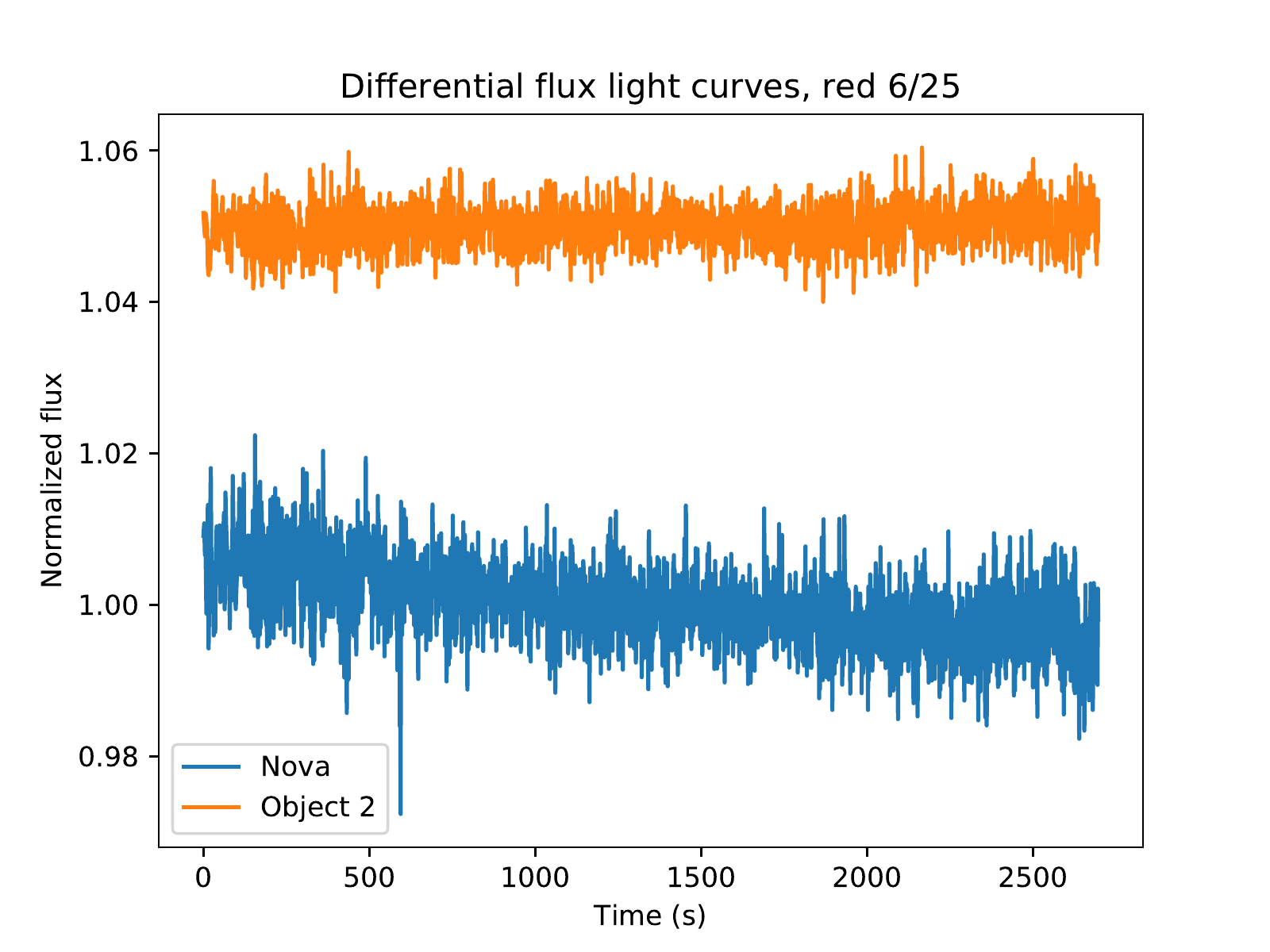}

\caption{Red differential light curves of the nova and the brightest reference star, taken on June 25th, 2017. The light curves have been normalized to a mean of 1 and offset by a constant. This plot shows the differential data before it has been linearly de-trended for analysis.}
\end{center}
\end{figure}

After differential normalization, we linearly de-trended the light curves, using a least-squares linear fit and model subtraction. The light curves produced from our first night of data, both uncalibrated and normalized, are shown in Figure 4. There is still some residual variation due to changing airmass and the variation in color between the target and reference stars, on timescales longer than an hour. However, variations on these timescales do not affect our periodicity search, and the associated frequency peaks in power are excluded from our analysis.

\subsection{Noise}

We initially assess the quality of our differential photometry by comparing the measured noise in the resulting light curves with theoretical expectations. Both read noise and dark current are essentially negligible in each frame relative to the photon noise (of both the target and references) and scintillation noise. Whereas photon noise is simply proportional to $\sqrt{N}$, where $N$ is the number of electron counts from the source, scintillation noise is affected by the properties of the observing telescope and observation parameters. \cite{Scintillation} provides an estimate of a point source's fractional amplitude fluctuation $\sigma_Y$:

\begin{equation}
\sigma_Y^2 = 10^{-5}D^{-4/3}t^{-1}(\textrm{cos}\gamma)^{-3}e^{-2h_{obs}/H}
\end{equation}

where $D$ is the telescope diameter, $t$ is the observation exposure time, $\gamma$ is the zenith distance, $h_{obs}$ is the observatory altitude, and $H$ is the atmospheric turbulence scale height, roughly 8000 m. All parameters are in SI units. \cite{Scintillation} also note that this approximation actually underestimates the mean scintillation noise at several observatories around the world; the authors provide modified equations to more accurately estimate scintillation noise. However, these more accurate methods require measurements of the atmospheric turbulence profile as a function of altitude at the relevant observatory, preferably taken on the same night as the photometric observations. The MASS-DIMM robotic system at Palomar is capable of monitoring atmospheric turbulence, but has been inactive for several years \citep{MASS-DIMM}.

For the Palomar 200-inch telescope, at an exposure time of 1.1 s and airmass 1.68 (the starting airmass of ASASSN-17hx for our 6/25 observations), $\sigma_Y = 1.80\times10^{-3}$. This value is within an order of magnitude of the true fractional noise $\sigma/\mu$ measured on most of our observing nights, as recorded in Table 2. Airmass of our observations on all nights ranges between 1 and 2.68, meaning that our estimate of $\sigma_Y$ ranges from $8.25\times10^{-4}$ to $3.62\times10^{-3}$. This is roughly the same range spanned by the theoretical Poisson noise calculated on all nights. We therefore conclude that scintillation noise dominates for the timescales of interests, limiting our 1~s cadence data to a few mmag precision (Table 2), which is further reduced to a $\sim$mmag precision when binning to longer timescales (Fig. 5).

\begin{figure}
\begin{center}
\includegraphics[width = 0.49\textwidth]
{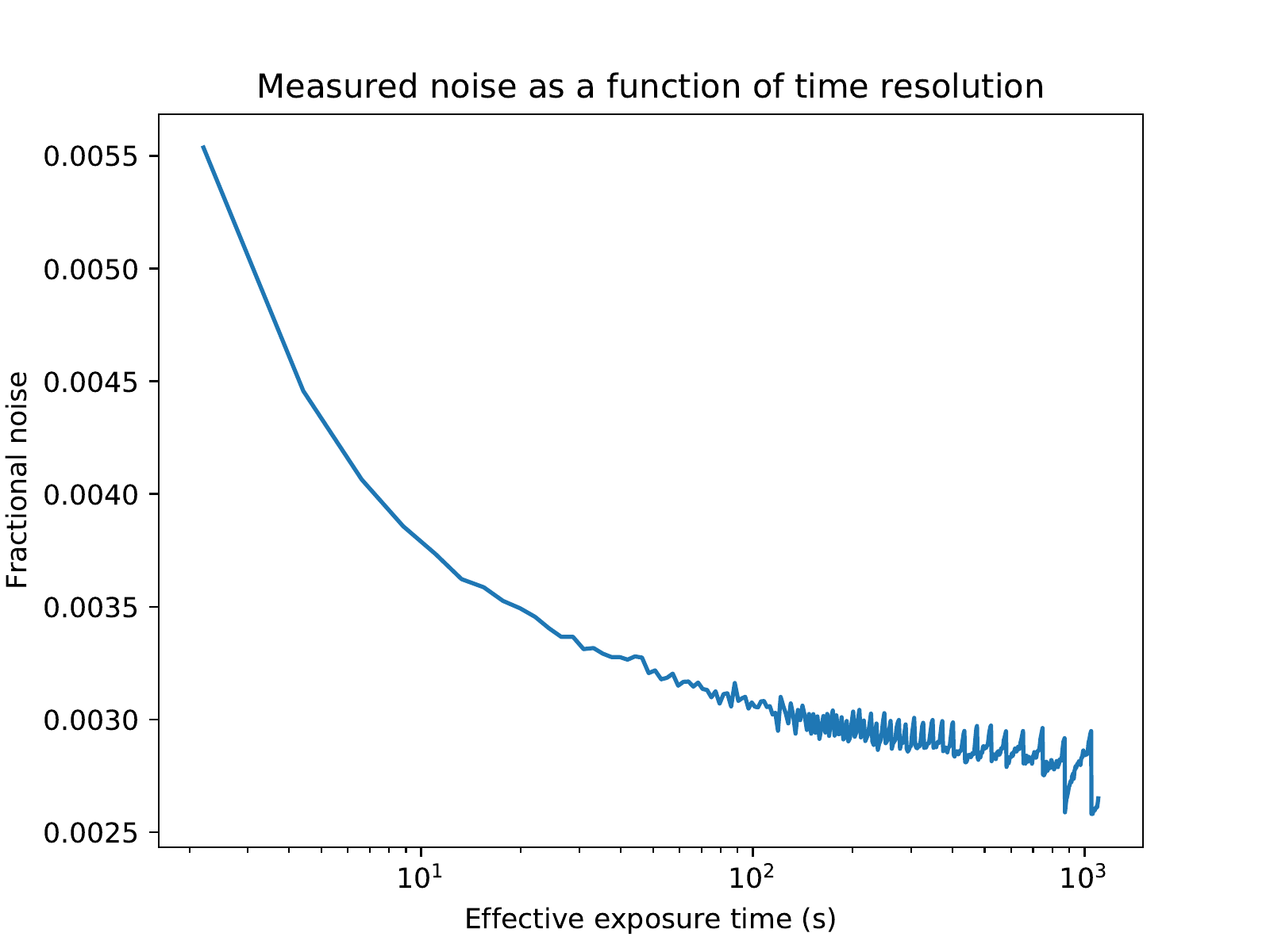}
\caption{Measured noise in the red 6/25 data as a function of binned time resolution. Fractional noise asymptotes around $0.3\%$.}
\end{center}
\end{figure}

\begin{table*}[t]
 \centering
  \begin{tabular}{ c || c | c | c | c | c }
  	Dataset & $\mu$ & True $\sigma$ & Poisson $\sqrt{\mu}$  & $\sigma/\mu$ & $1/\sqrt{\mu}$ \\ \hline \hline
    Blue 6/25 & $3.73\times10^5$ & $2.81\times10^3$ & $6.11\times10^2$ & $7.53\times10^{-3}$ & $1.64\times10^{-3}$ \\
    Red 6/25 & $8.52\times10^5$ & $4.72\times10^3$ & $9.23\times10^2$ & $5.54\times10^{-3}$ & $1.08\times10^{-3}$ \\
    Blue 6/26 & $2.60\times10^5$ & $6.55\times10^3$ & $5.10\times10^2$ & $2.52\times10^{-2}$ & $1.96\times10^{-3}$ \\
    Red 6/26 & $1.12\times10^6$ & $9.25\times10^3$ & $1.06\times10^3$ & $8.26\times10^{-3}$ & $9.43\times10^{-4}$ \\
    Blue 7/22 & $9.52\times10^5$ & $7.91\times10^3$ & $9.76\times10^2$ & $8.31\times10^{-3}$ & $1.02\times10^{-3}$ \\
    Red 7/22 & $1.44\times10^6$ & $1.26\times10^4$ & $1.20\times10^3$ & $8.75\times10^{-3}$ & $8.33\times10^{-4}$ \\
    Blue 7/23 & $2.17\times10^6$ & $1.58\times10^4$ & $1.47\times10^3$ & $7.28\times10^{-3}$ & $6.80\times10^{-4}$ \\
    Red 7/23 & $1.26\times10^6$ & $1.62\times10^4$ & $1.12\times10^3$ & $1.29\times10^{-2}$ & $8.93\times10^{-4}$
  \end{tabular}
  \caption{Nova mean brightness, standard deviation in brightness, mean Poisson counting noise, and fractional variations for all observations, in instrumental photo-electron units. It should be noted that these measures are all computed for the post-differential photometry light curves. We did this to avoid including cloud cover in our noise measurements, but at the cost of adding the Poisson noise of our reference stars in quadrature to the nova Poisson noise.}
  \label{tab:2}
\end{table*}

\subsection{Search for periodicity}

We searched for periodicity in our nova light curves by calculating Lomb-Scargle periodograms \citep{LSprime, LombScargle, LombScargle1976, Scargle82} for each night of data, and used a bootstrap method to measure the statistical significance of periodic signals. The Lomb-Scargle periodogram is an algorithm that essentially computes the Fast Fourier Transform of non-uniformly sampled time-series data. This takes into account the windowing effect caused by time delays between the FITS cubes in which CHIMERA data is saved, as well as some longer gaps in our data. Figures 6 through 9 show the computed power spectra of the nova and reference star light curves, in both color filters and on all nights.

In order to evaluate the statistical significance of peaks in a periodogram, we used a bootstrap method to generate a roughly normal distribution of peak periodicity in our light curve data. We then used this distribution to measure the probability that a periodicity peak could arise out of random noise. To obtain this distribution, we iterated 1000 times over each normalized nova light curve. During each iteration over a time series (replaced 'light curve') containing N data points, we shuffled the data and randomly sampled with replacement N times to construct a new light curve.

We then generated a Lomb-Scargle periodogram for that randomly generated light curve, recorded the periodogram’s peak value, and appended it to our periodicity-peak distribution. We used this distribution to compute the periodicity peak strengths that correspond to 5 percent and 0.3 percent likelihood that the peak arose from random noise, and included these measures in plots of the true data's Lomb-Scargle periodograms, seen in Figures 6 through 9.

\begin{figure*}[h]
\begin{center}
\includegraphics[width = 0.49\textwidth]{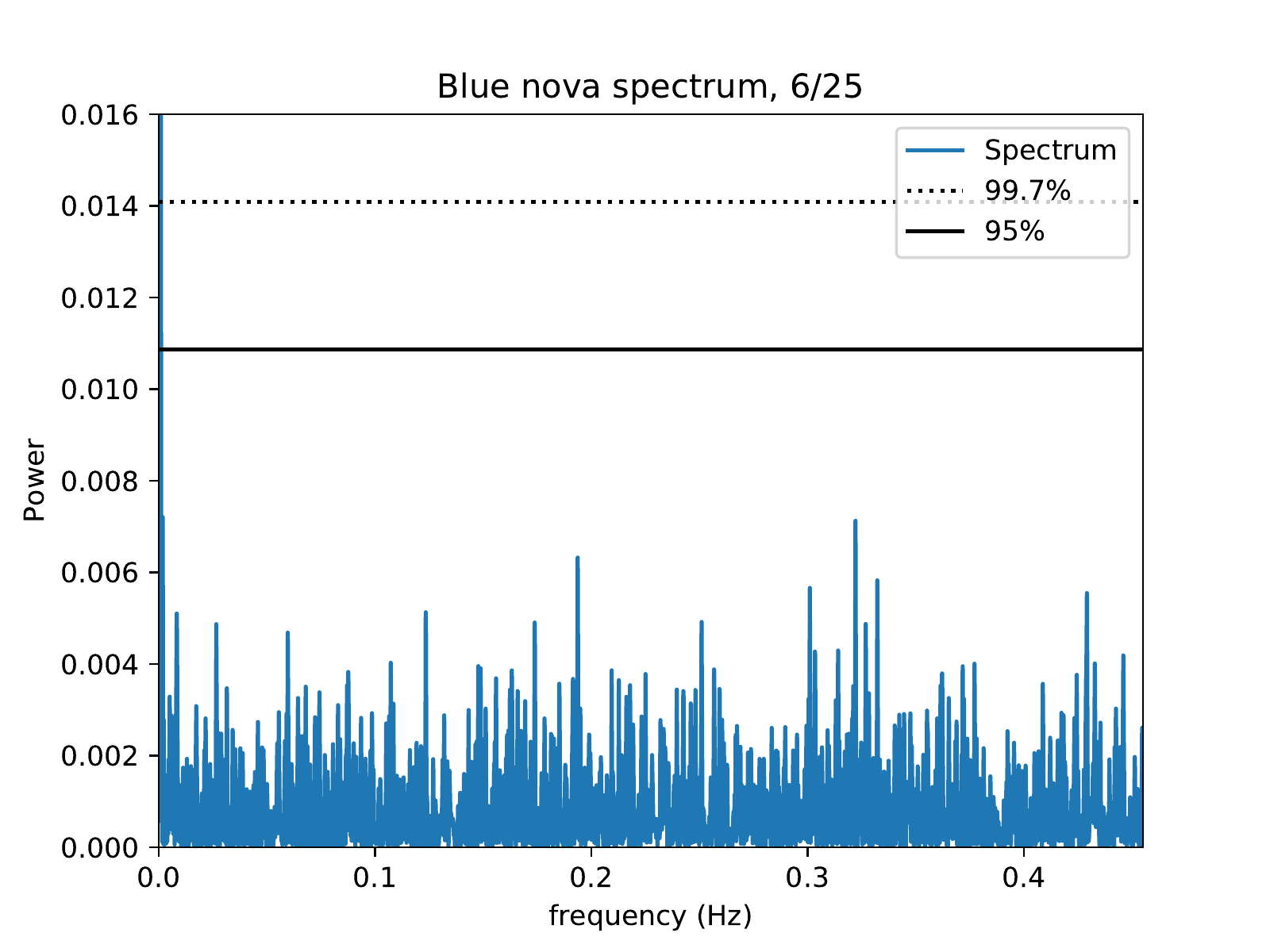}
\includegraphics[width = 0.49\textwidth]{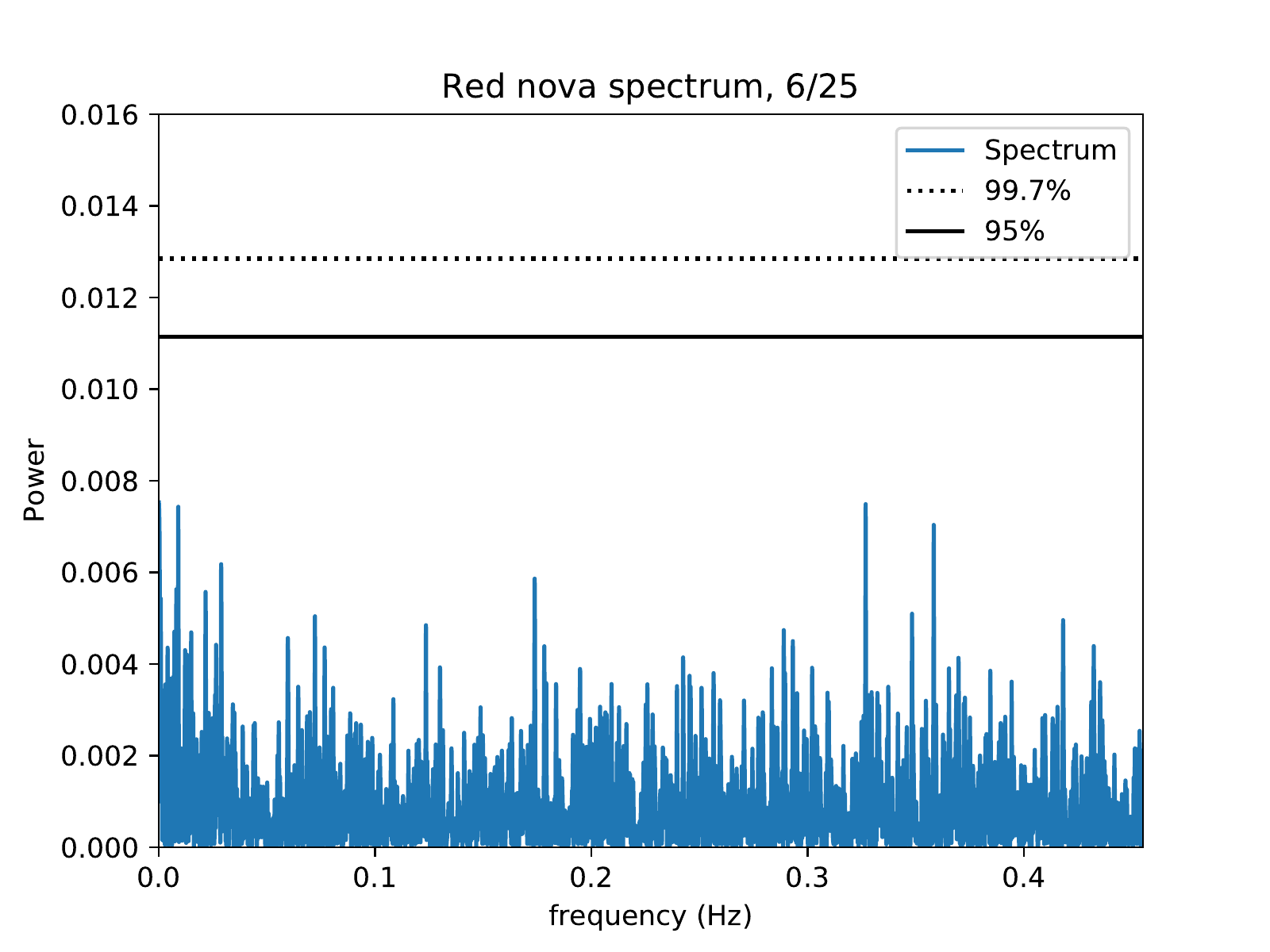}\\
\includegraphics[width = 0.49\textwidth]{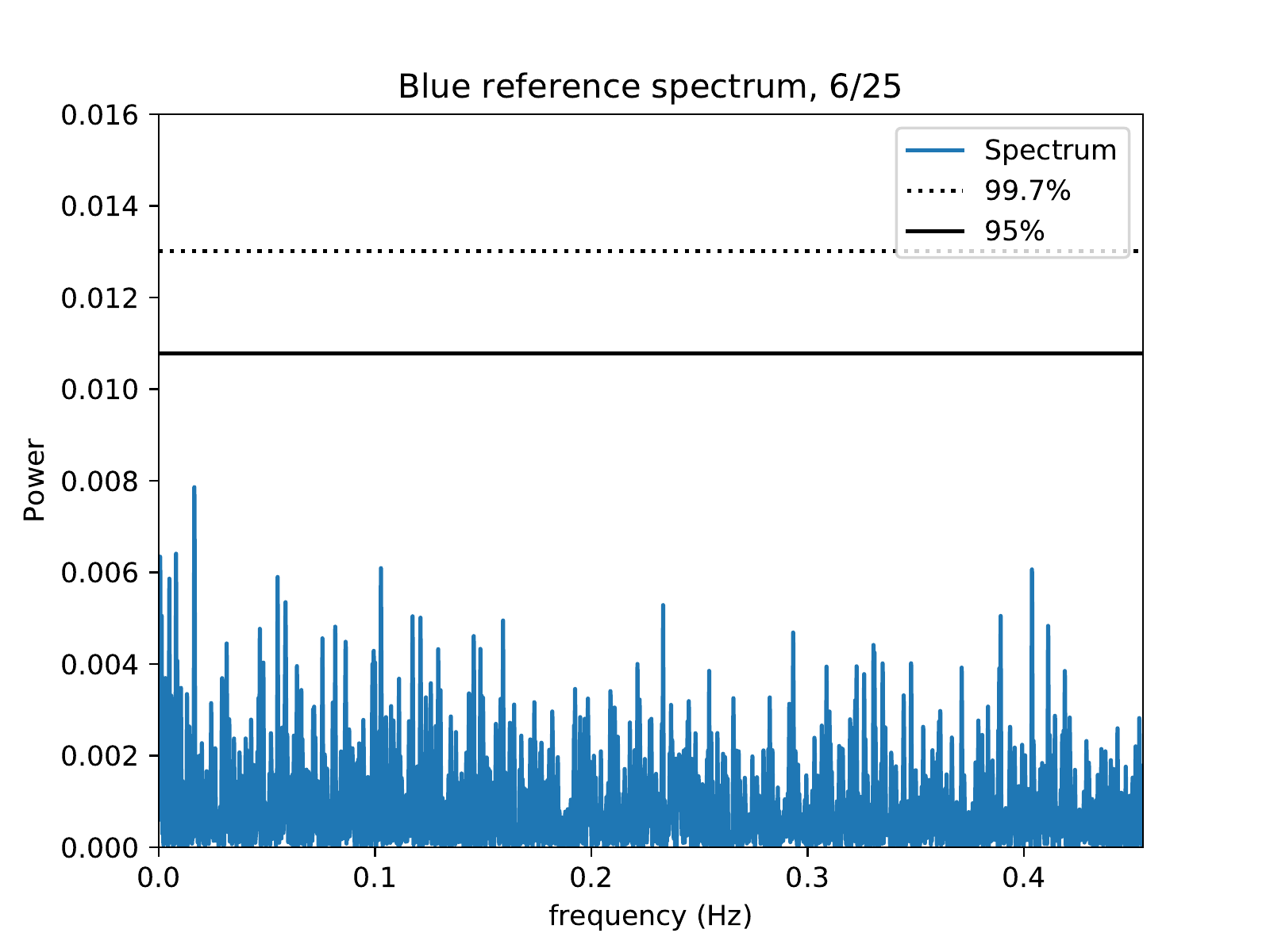}
\includegraphics[width = 0.49\textwidth]{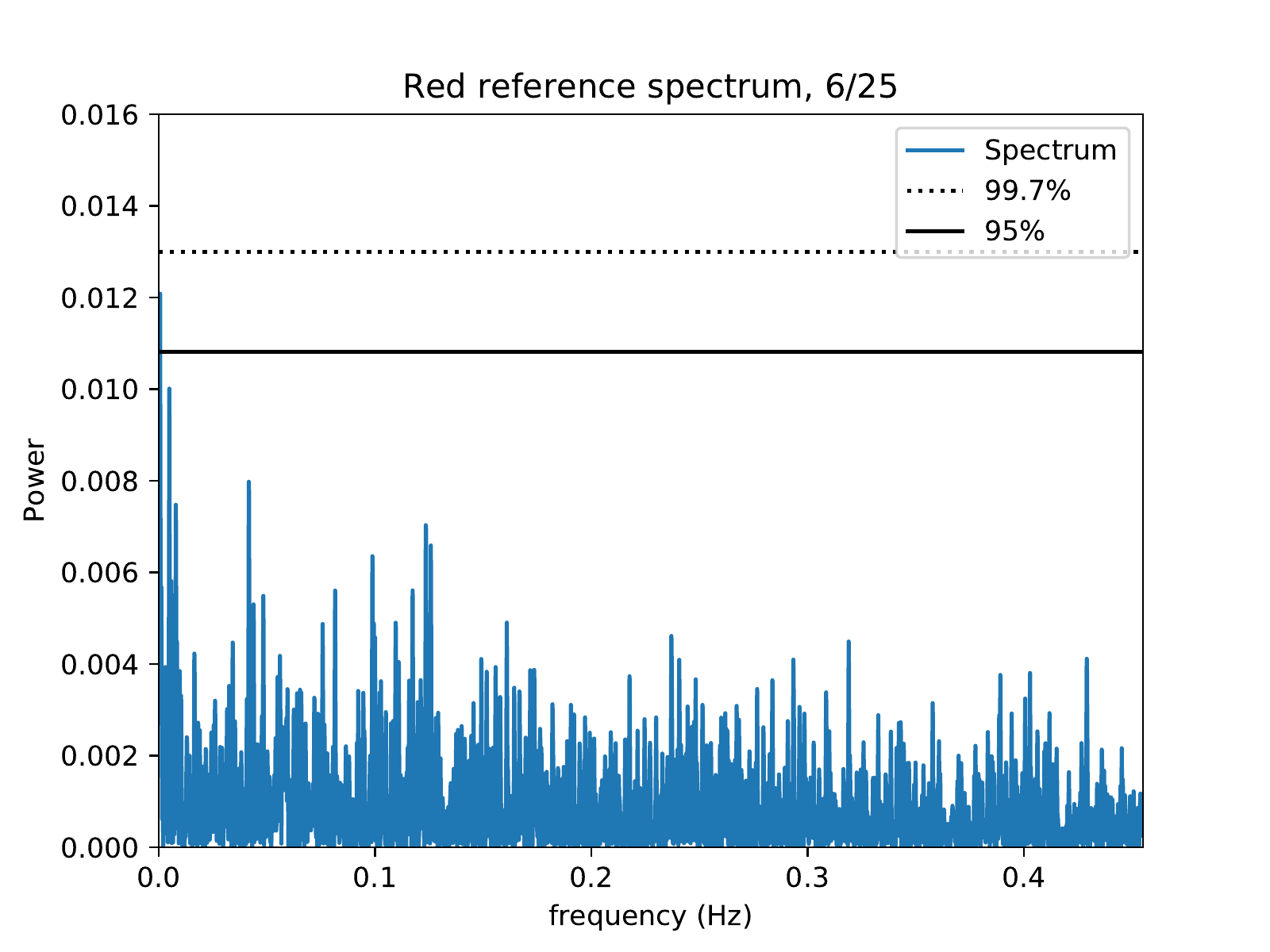}\\

\caption{Lomb-Scargle periodograms for ASASSN-17hx and the brightest reference star in the field of view, from our 6/25 data. There are no periodic signal spikes stronger than $2\sigma$ significance, determined by bootstrapped limits as described in the methods section, with the exception of a spike approaching zero frequency in the blue nova spectrum and the red reference spectrum. These are possibly due to residual, constant offsets, leftover from the linear fit subtraction. Another possible explanation is a nonlinear trend due to changing airmass.}
\end{center}
\end{figure*}

\begin{figure*}[h]
\begin{center}
\includegraphics[width = 0.49\textwidth]{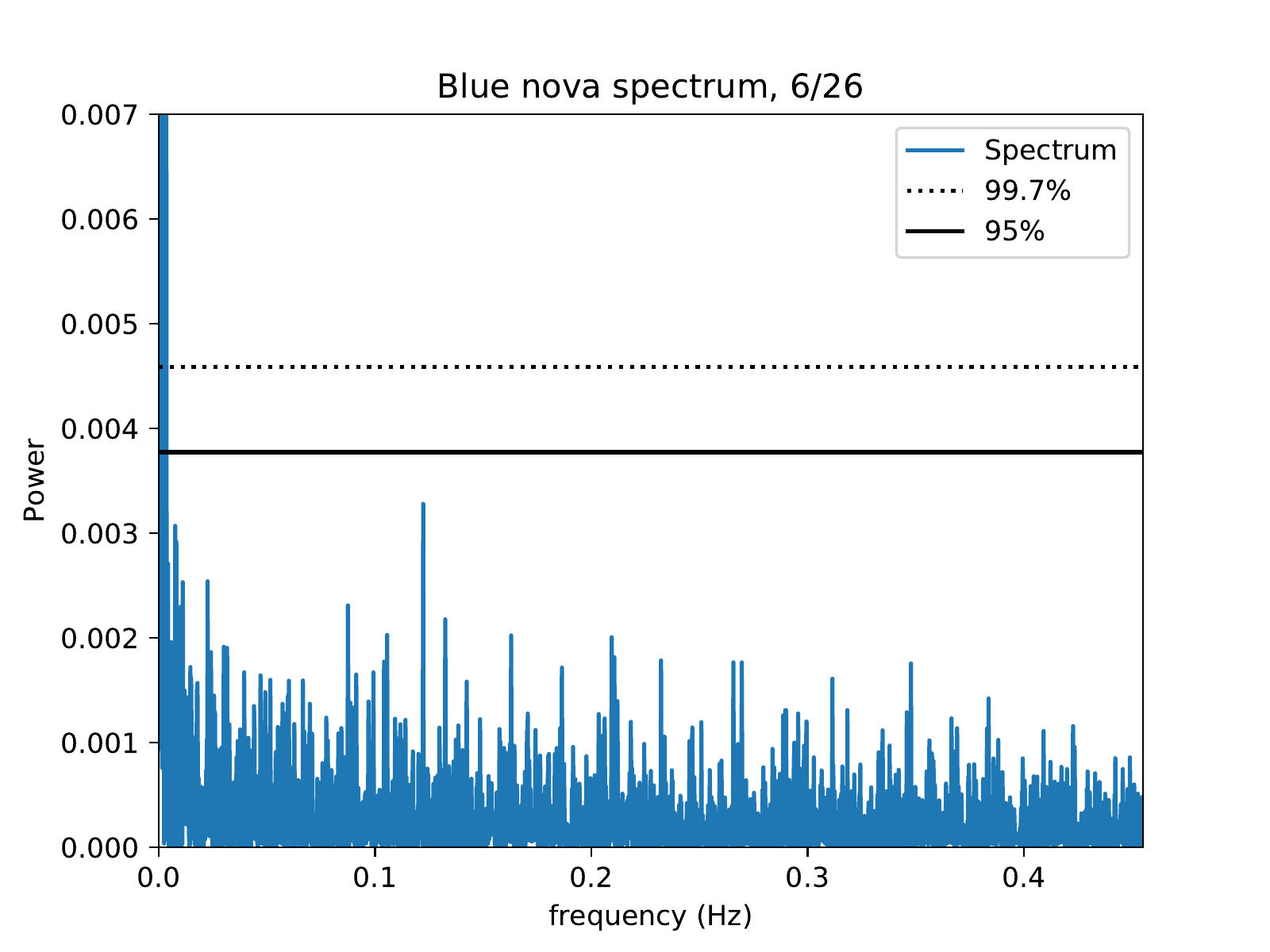}
\includegraphics[width = 0.49\textwidth]{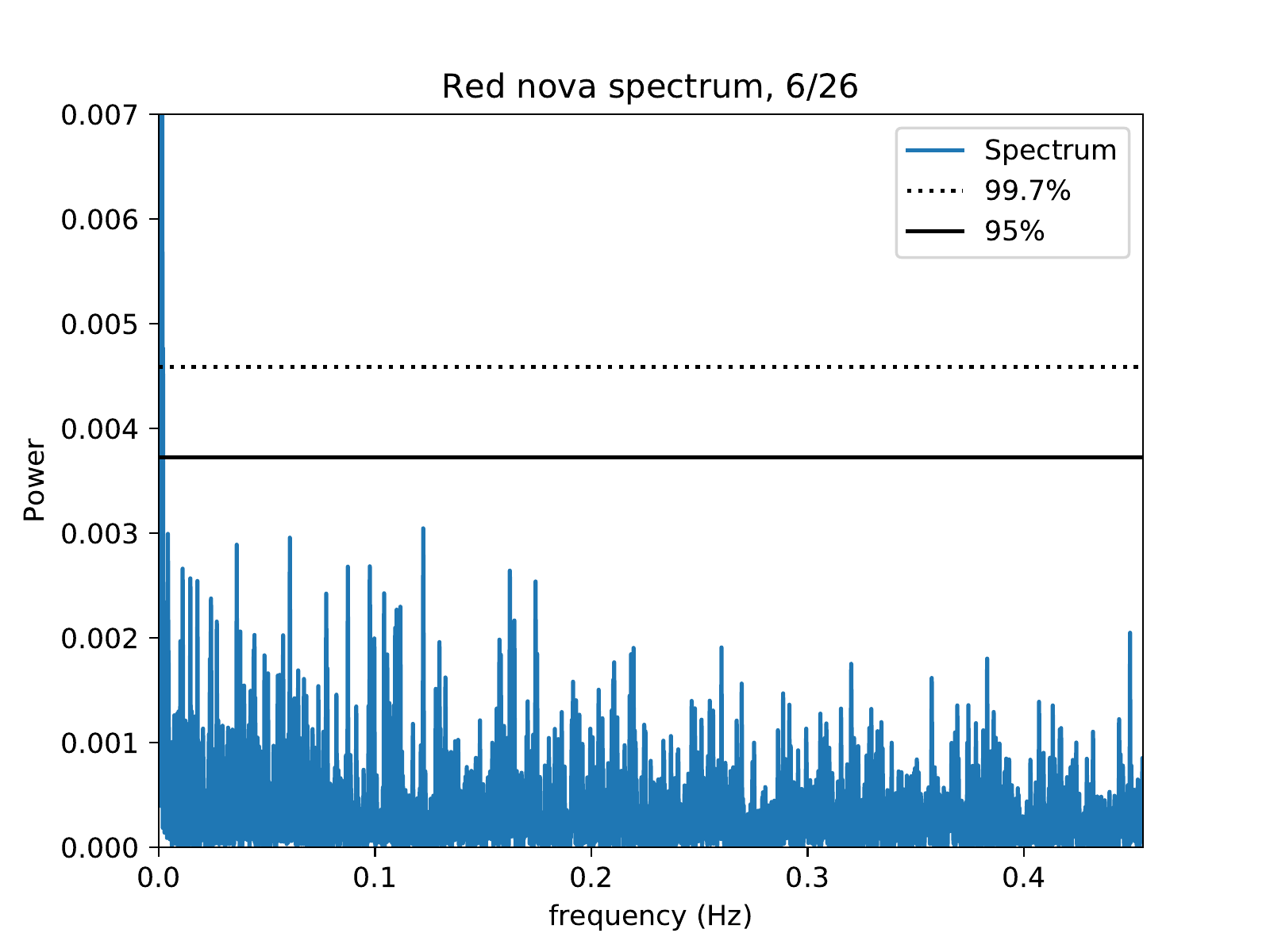}\\
\includegraphics[width = 0.49\textwidth]{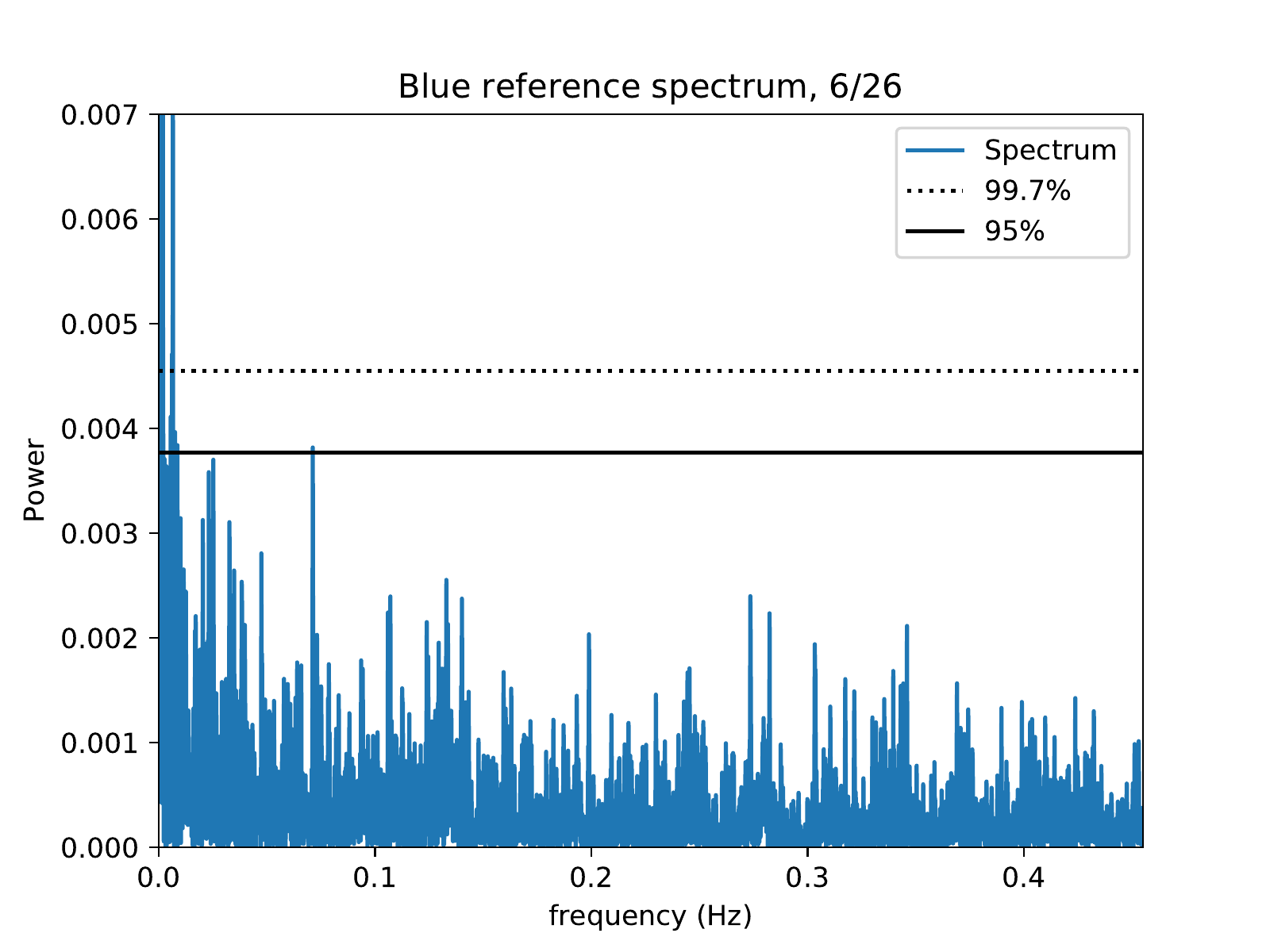}
\includegraphics[width = 0.49\textwidth]{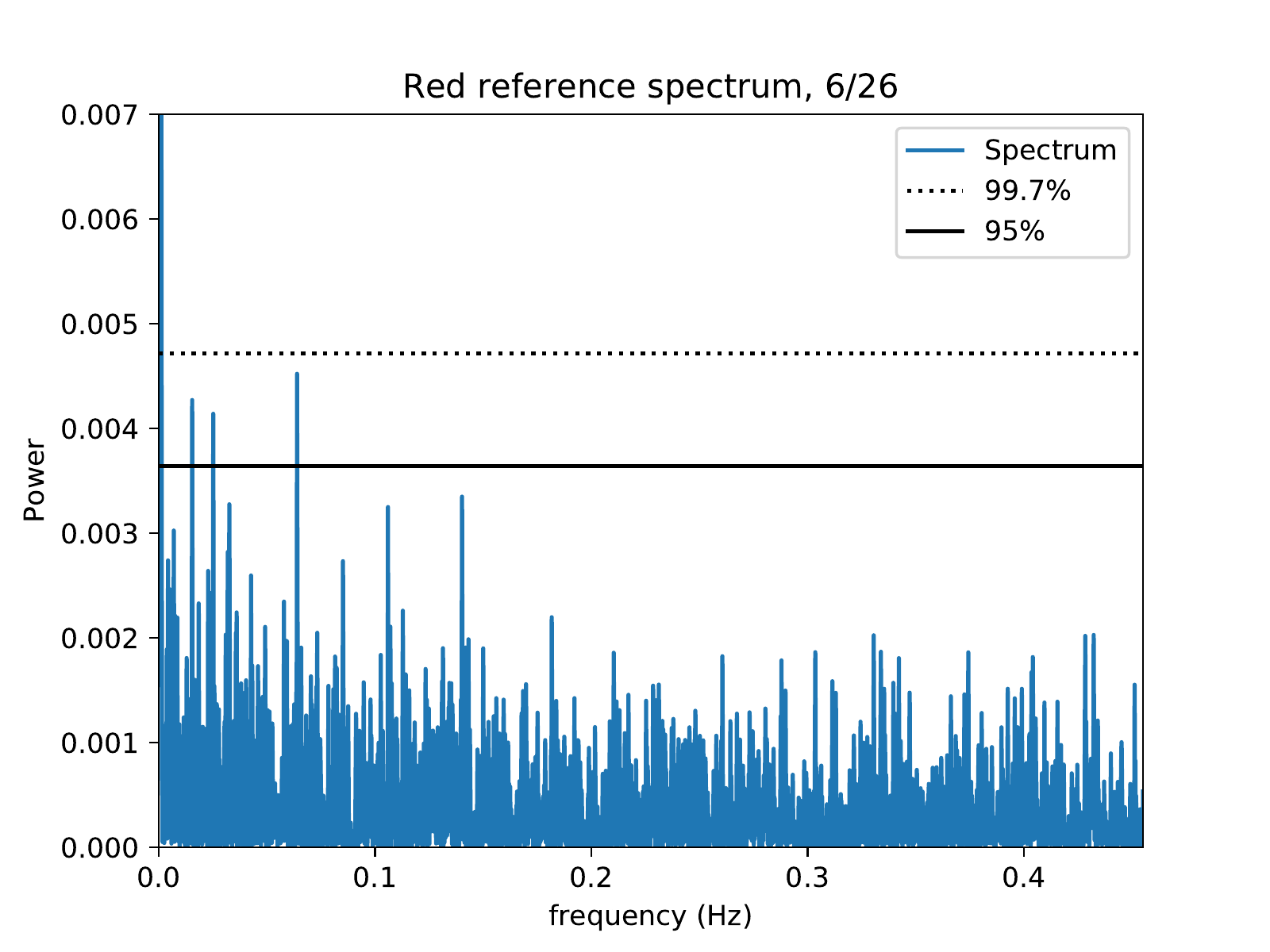}\\

\caption{Lomb-Scargle periodograms from our 6/26 data. On this night, there are signals approaching zero frequency in all of the periodograms, again due to linear fit residuals, or possibly a nonlinear trend due to changing airmass. Aside from the zero-frequency peaks, there are no statistically significant peaks in the blue and red nova periodograms.}
\end{center}
\end{figure*}

\begin{figure*}[h]
\begin{center}
\includegraphics[width = 0.49\textwidth]{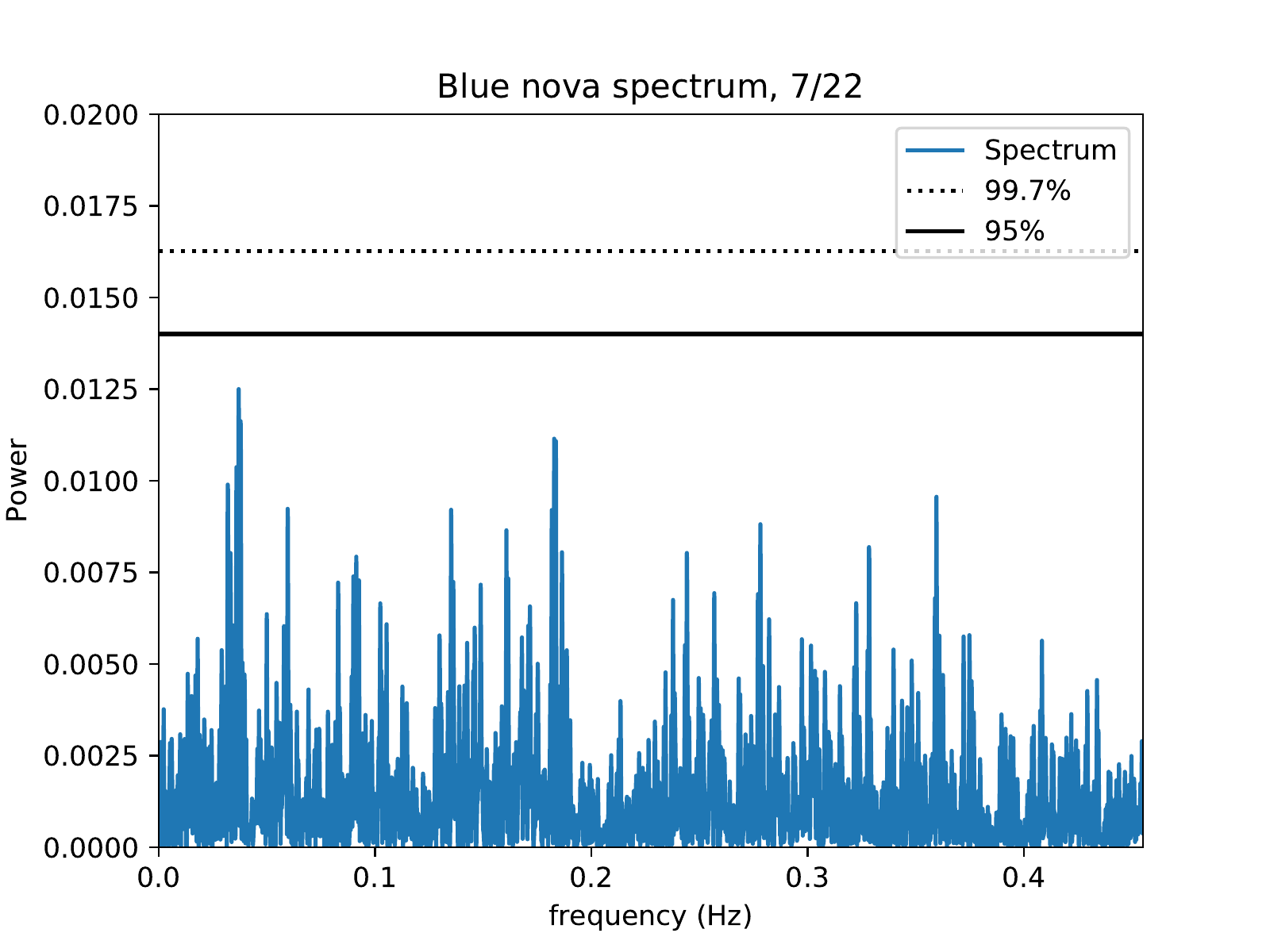}
\includegraphics[width = 0.49\textwidth]{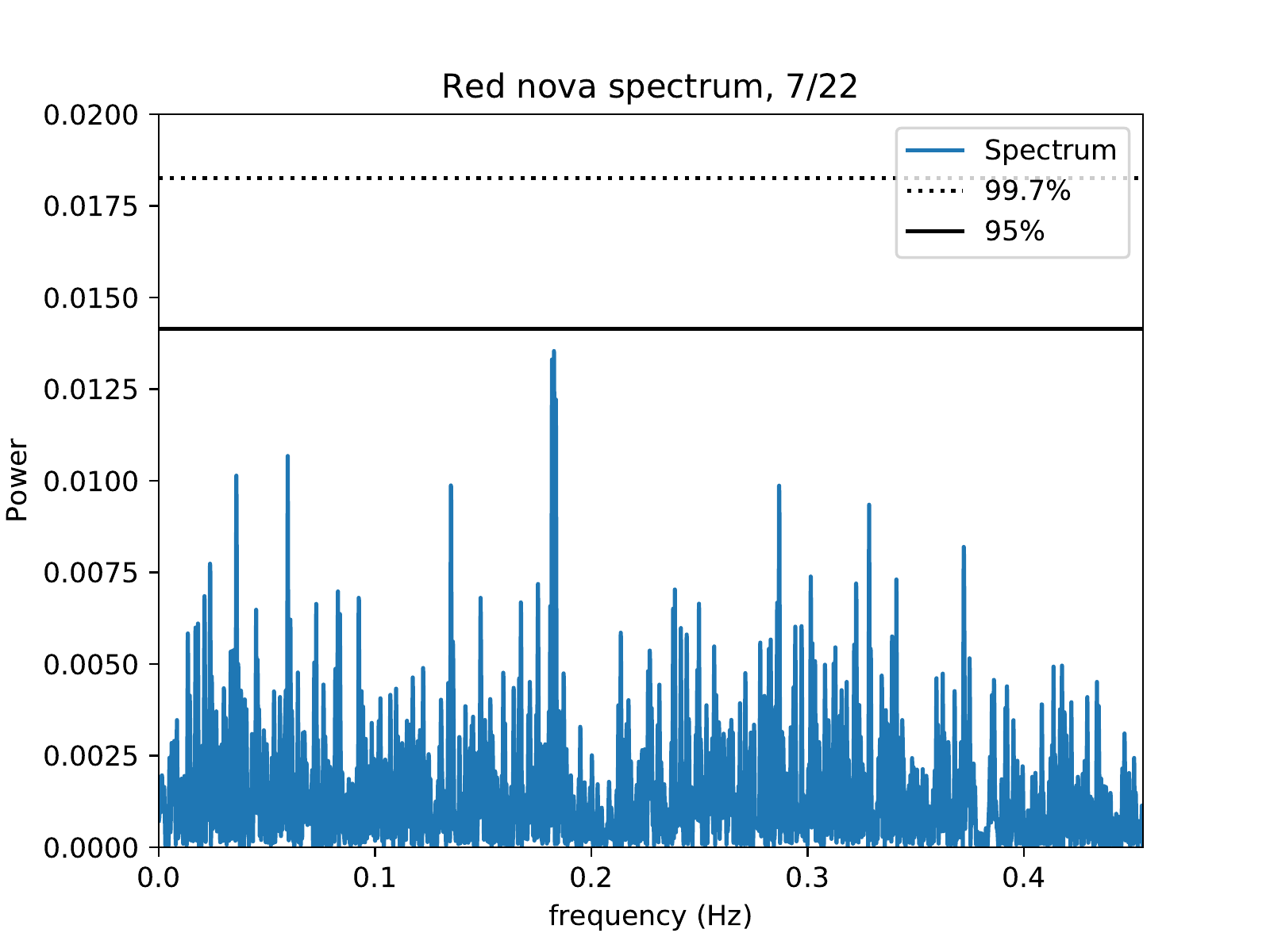}\\
\includegraphics[width = 0.49\textwidth]{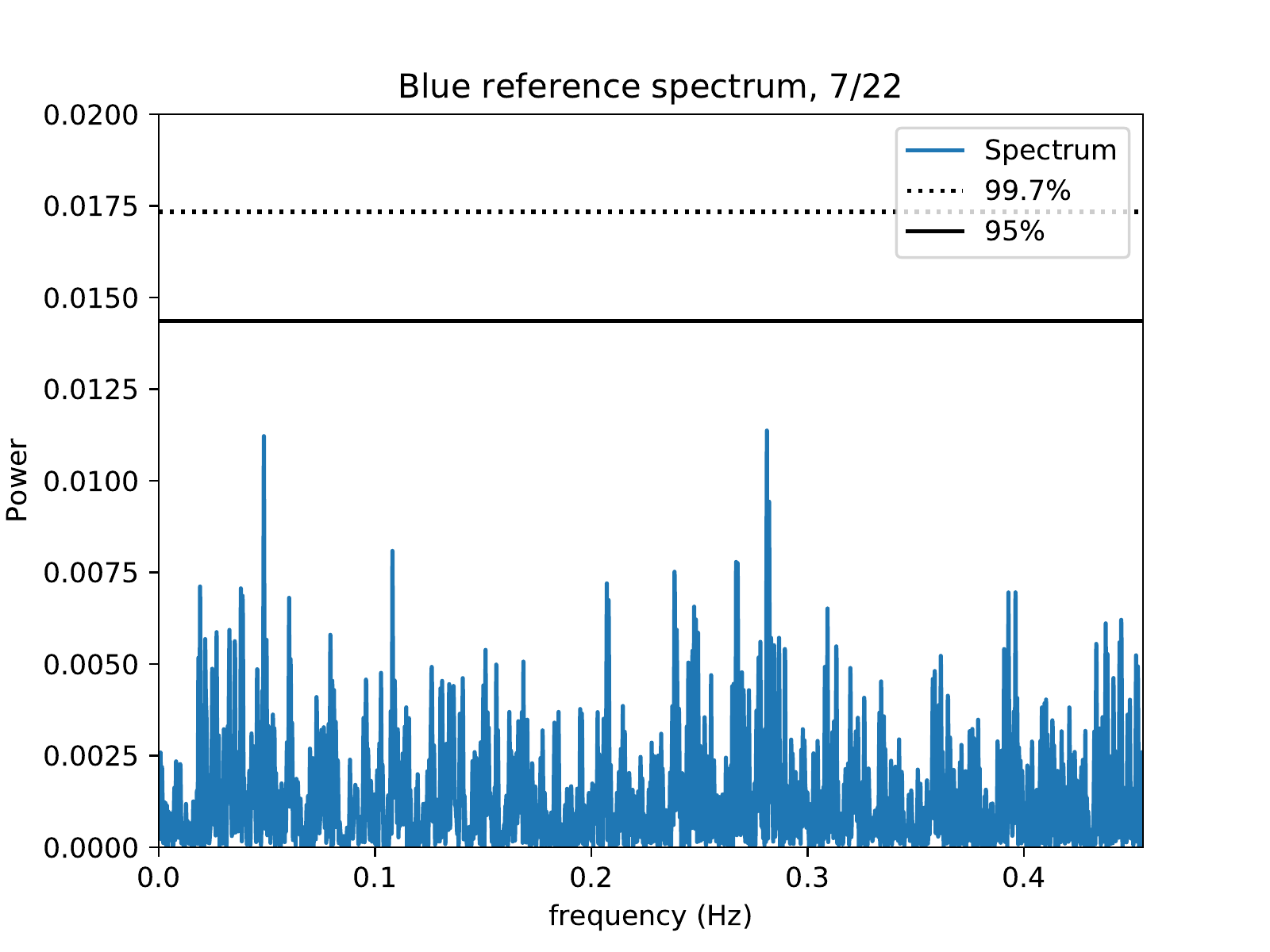}
\includegraphics[width = 0.49\textwidth]{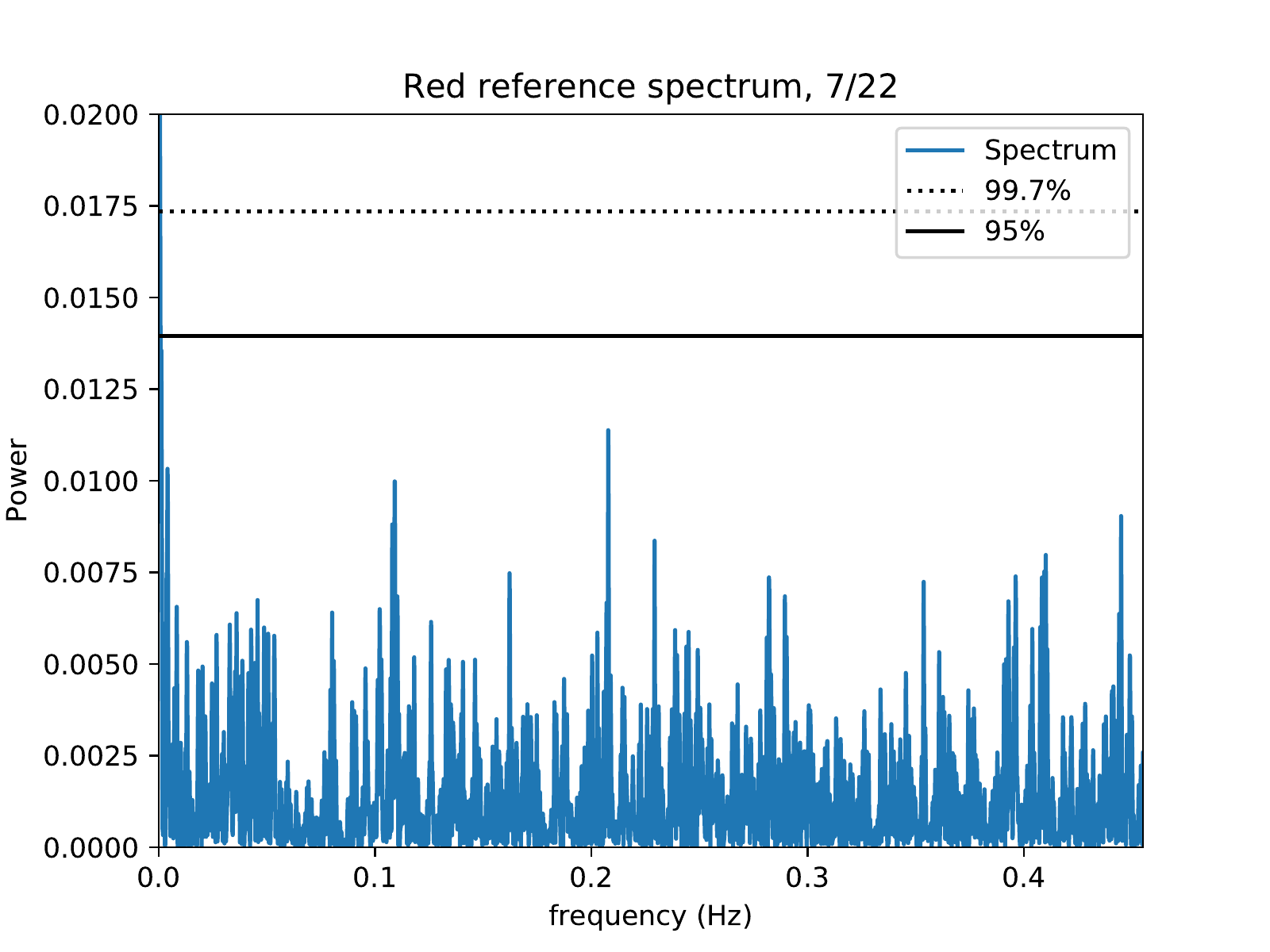}\\

\caption{Lomb-Scargle periodograms from our 7/22 data. As seen in Figure 4, which shows differential curves from this night, there is significant residual cloud cover in the data. We window out the contaminated sections of data before conducting our periodicity analysis. Even after this, there is a signal on the red side with significance close to $2\sigma$, at roughly 0.19 Hz. As this signal is only seen on the red side, and does not surpass the $3\sigma$ threshold, it is more likely to be caused by subtle, residual cloud cover that our filtering did not remove, than to be caused by nova periodicity.}
\end{center}
\end{figure*}

\begin{figure*}[h]
\begin{center}
\includegraphics[width = 0.49\textwidth]{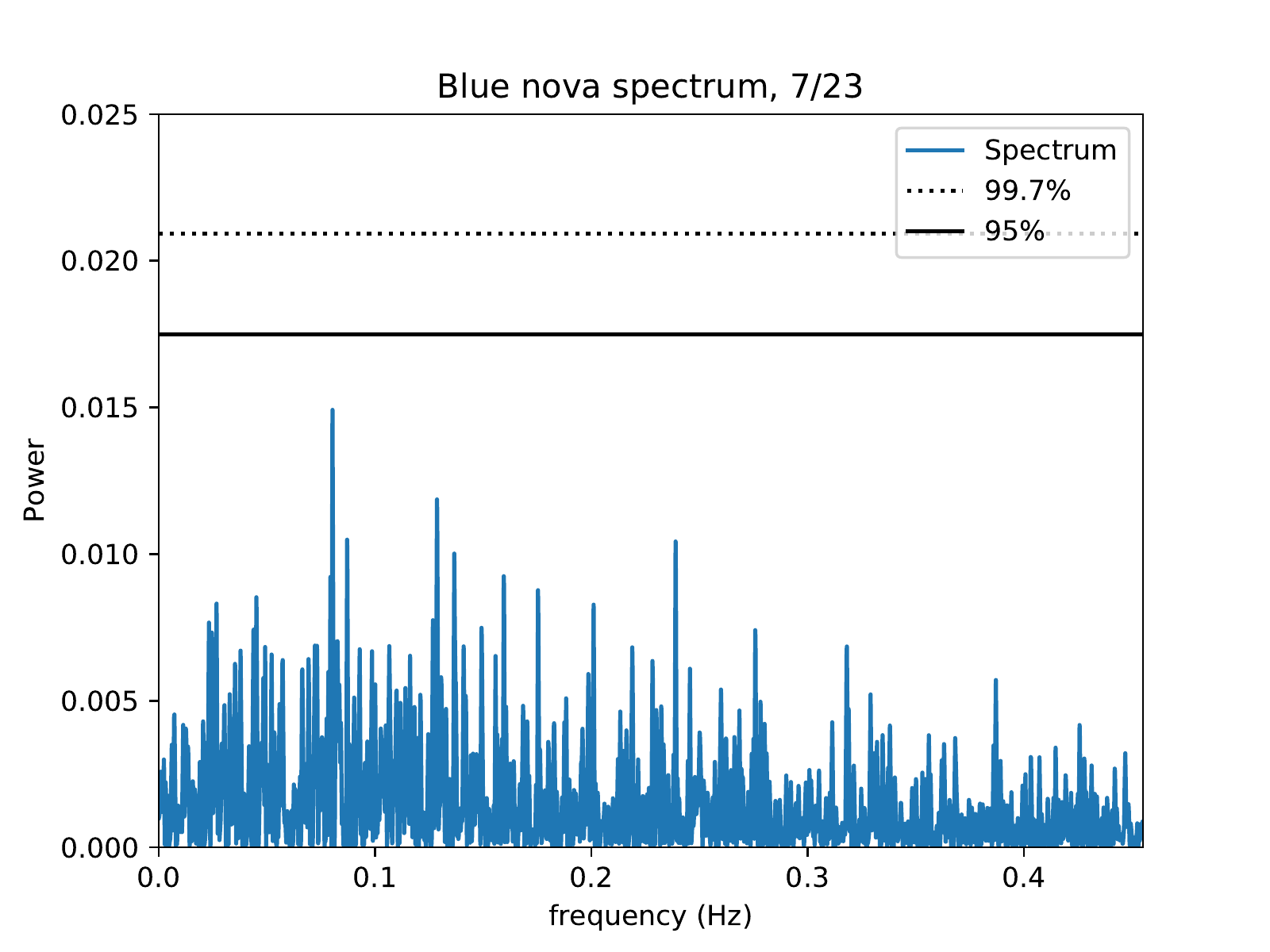}
\includegraphics[width = 0.49\textwidth]{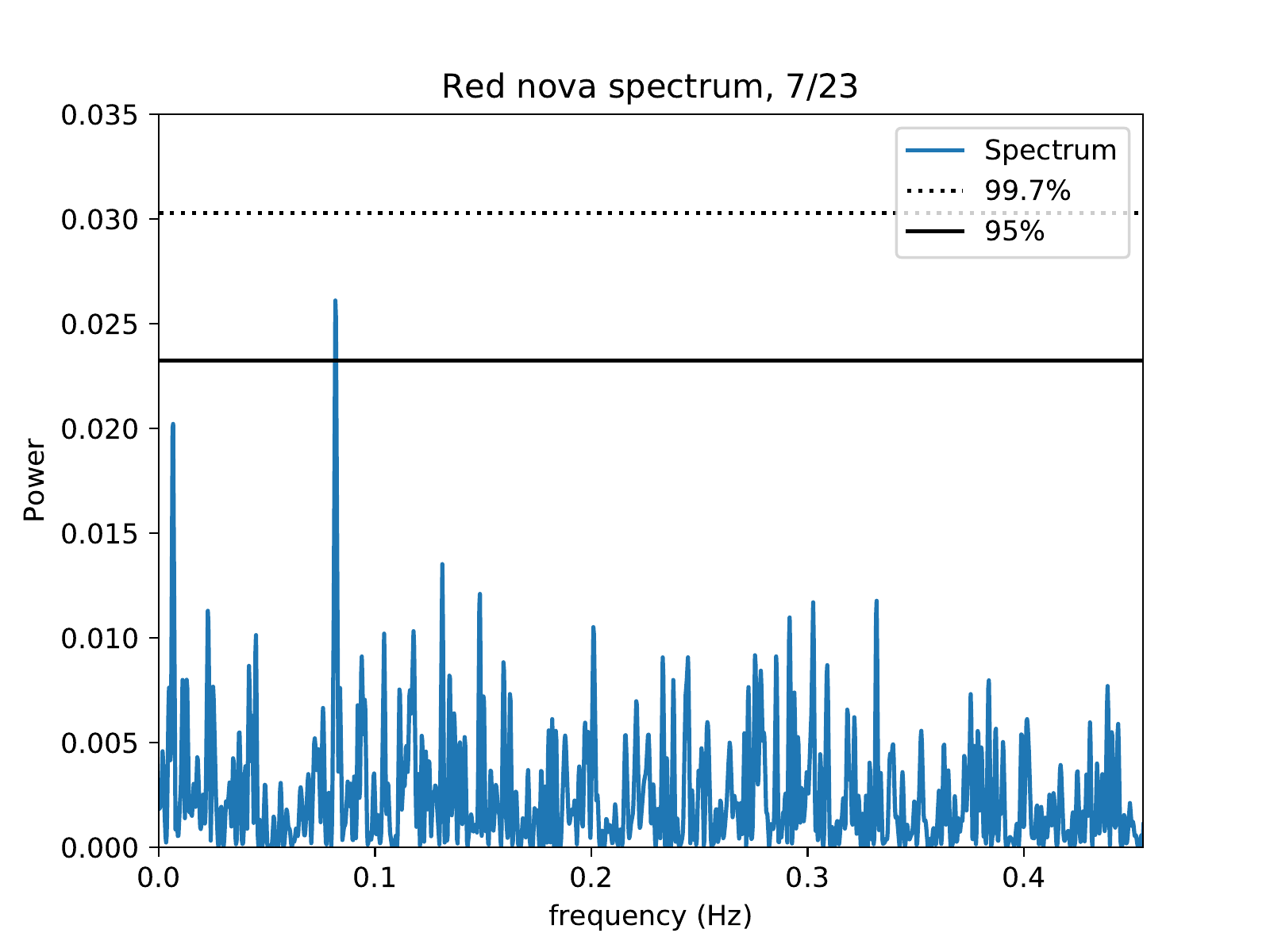}\\
\includegraphics[width = 0.49\textwidth]{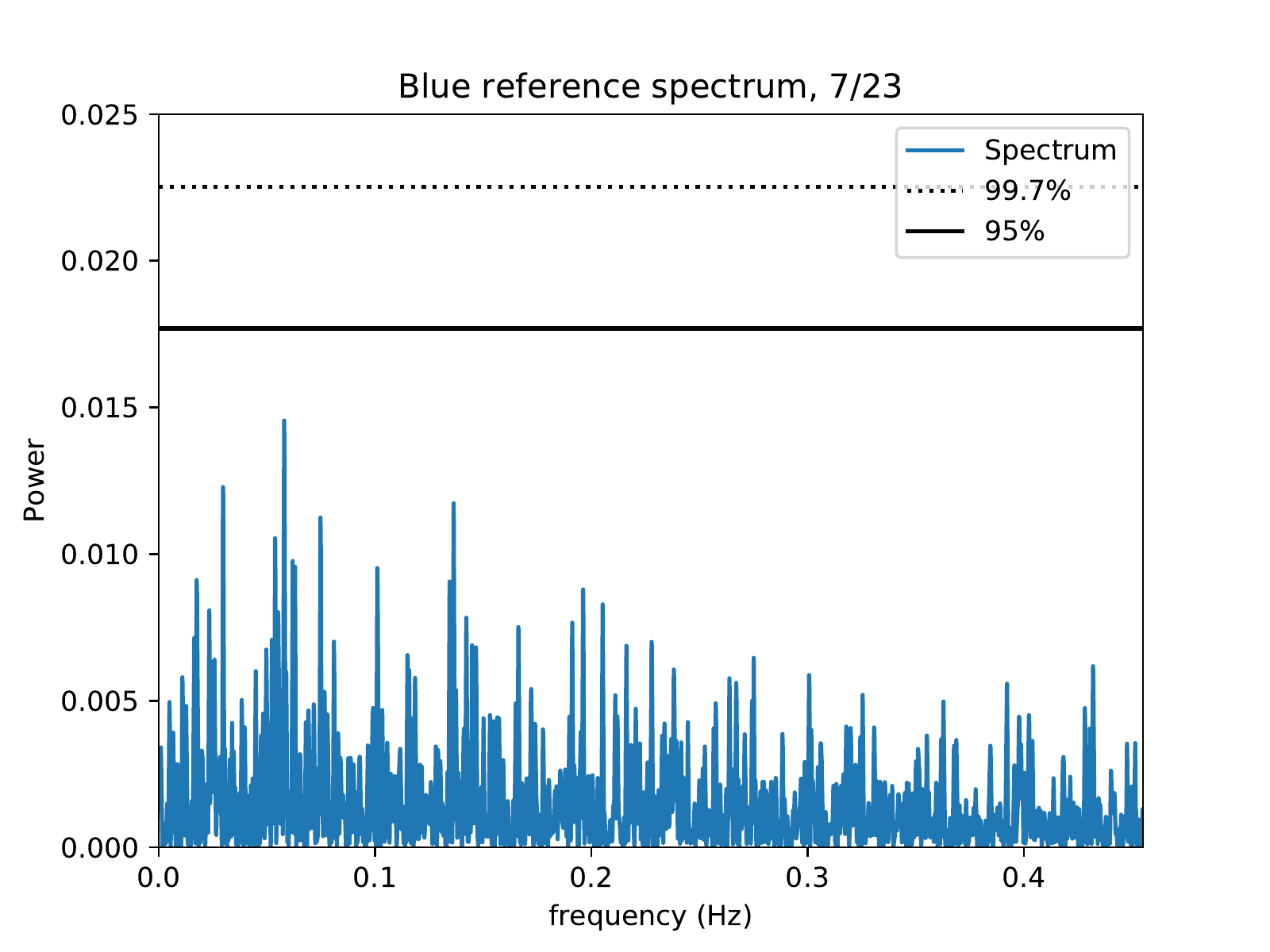}
\includegraphics[width = 0.49\textwidth]{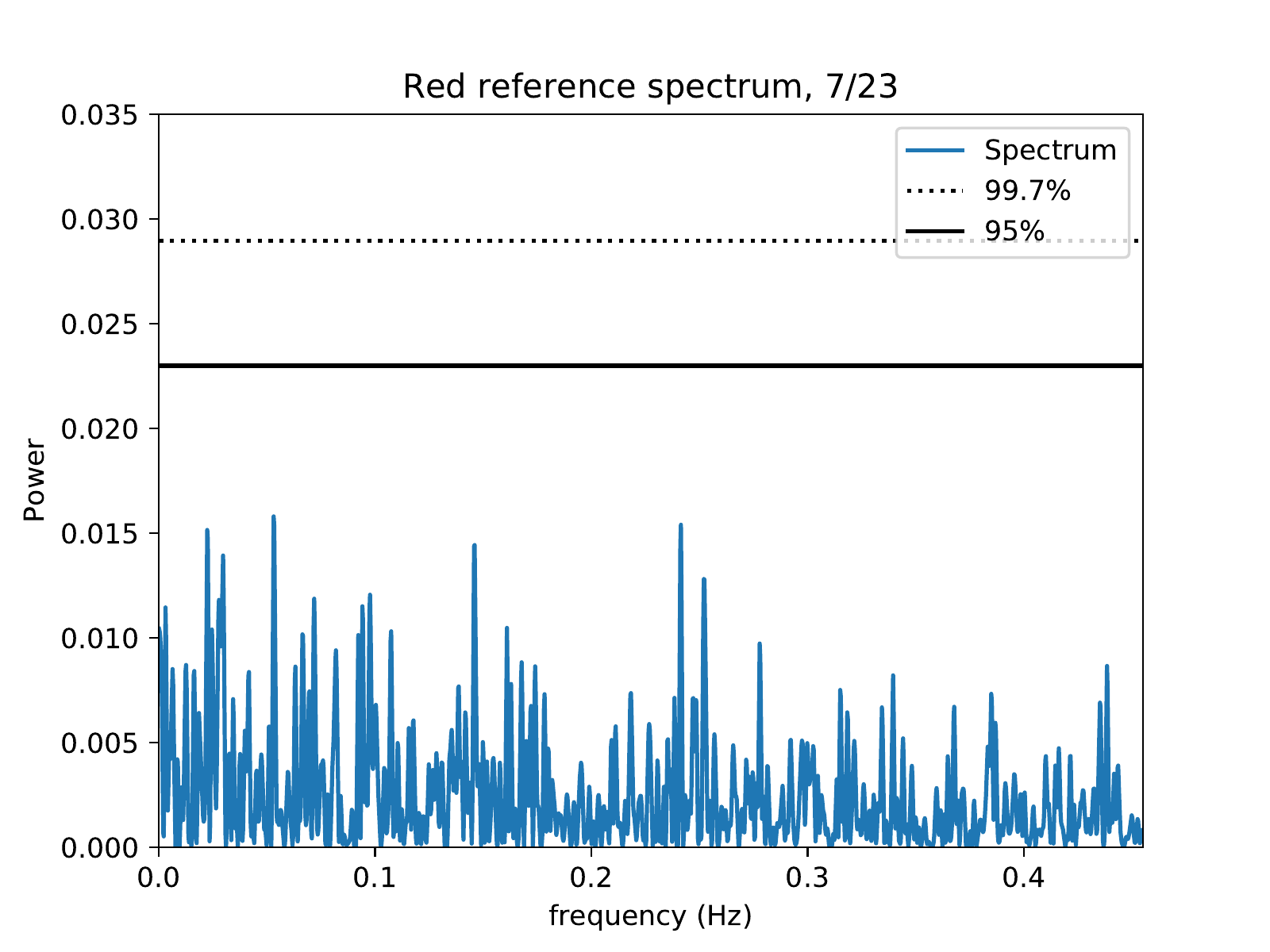}\\

\caption{Lomb-Scargle periodograms from our 7/23 data. On this night, the exposure time for the red side is shorter than the exposure time for the blue side, so we downsample the red light curves to have the same effective exposure time as that of the blue data. Therefore, the red spectra are less densely sampled in frequency space. There is one peak on the red side that has greater than $2\sigma$ significance, with a frequency around 0.08 Hz. The maximum peak of the blue spectrum has roughly the same frequency, but less than $2\sigma$ significance.}
\end{center}
\end{figure*}

\section{Results}

We do not find any signals of significance greater than $3\sigma$ in either channel, exploring timescales from 2 seconds to 16.7 minutes. As seen in Figure 9, we find a periodic signal with significance $2.42\sigma$ in the red nova light curve generated from our July 23rd data, close to 0.08 Hz. The maximum peak of the blue spectrum has roughly the same frequency, but less than $2\sigma$ significance.

We place constraints on the range of periodic signals detectable via CHIMERA observations by making an injection and recovery map using the normalized red 6/25 data and estimating a $3\sigma$ detection limit as a function of period. We iterated over a range of 64 period-amplitude pairs, injected a sinusoid with this pair of parameters into our light curve, and from this injected curve generated a Lomb-Scargle periodogram and bootstrapped detection limits. Finally, we measured the detection strength of our injected signal as the maximum peak of our periodogram in standard deviations, and plotted these strengths as a colormap, shown in Figure 10. 

Our colormap shows the $3\sigma$ limits as a function of period. This limit rules out the detection of periodic signals of fractional amplitude $>7.08\times10^{-4}$ on timescales of 2 seconds and fractional amplitude $>1.06\times10^{-3}$ on timescales of 10 minutes, and a range of amplitudes in between.

As another test of periodicity, we performed the Durbin-Watson test \citep{Durbin1, Durbin2} to measure autocorrelation in the linear-fit residuals of the differential nova light curves. Evidence of autocorrelation would imply the presence of correlated noise, which could be a periodic signal. However, we find no significant autocorrelation in any of the data.

\section{Conclusions}

We have observed the classical nova ASASSN-17hx to search for periodic signals that may arise from pulsations driven by convective motions within the expanding envelope. We rule out periodic signals of fractional amplitude $>7.08\times10^{-4}$ on timescales of 2 seconds and fractional amplitude $>1.06\times10^{-3}$  on timescales of 10 minutes, and a range of amplitudes in between. Additional theoretical work is required before these constraints can be interpreted within the framework of a physical model. Additional observations are planned for two more targets, probing to even shorter timescales.

\begin{figure}
\begin{center}
\includegraphics[width = \textwidth]{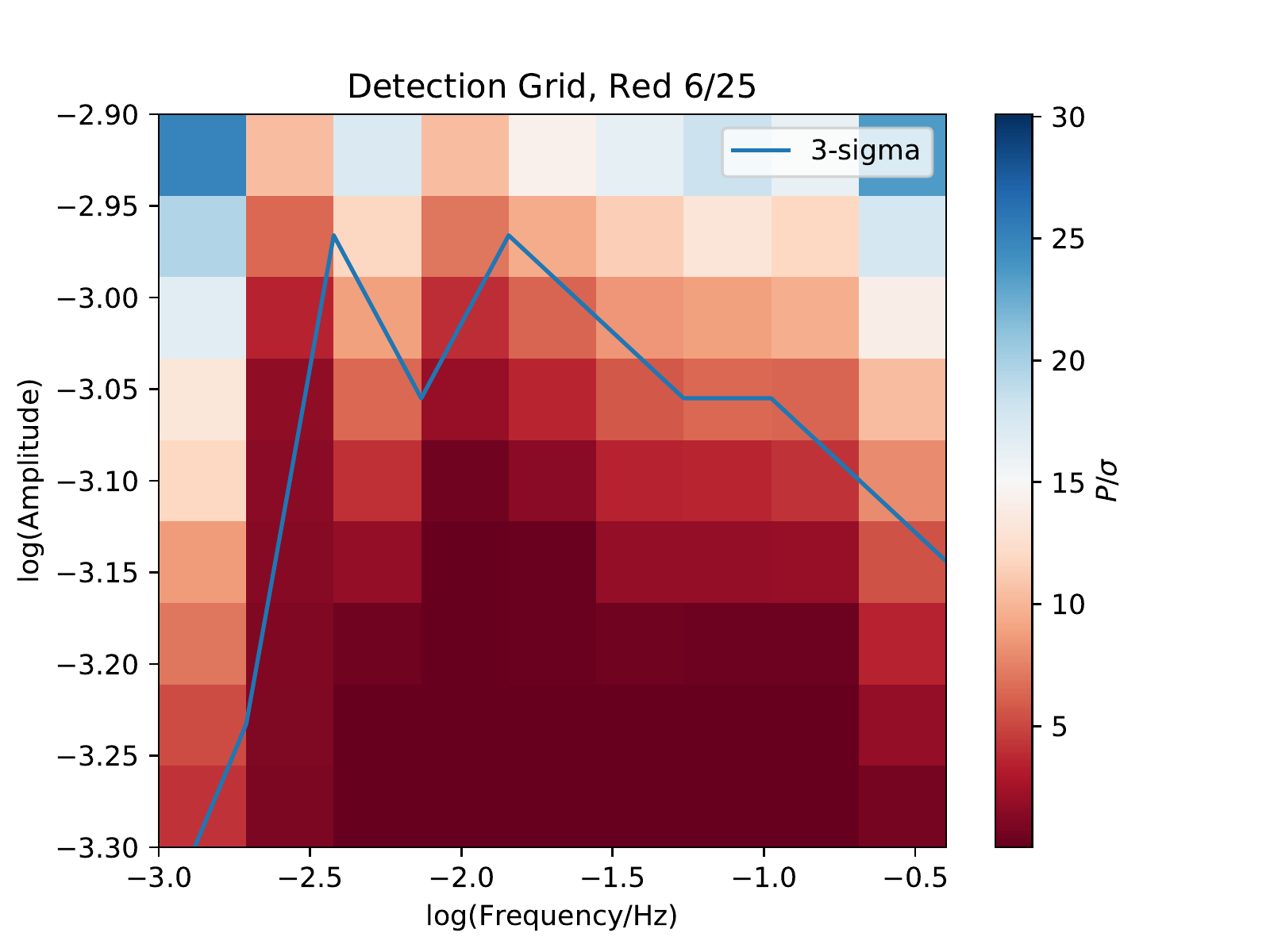}\\

\caption{Injection and recovery map. We normalized the red 6/25 differential light curve, by subtracting the linear proportionality of a linear fit and then dividing by the mean brightness. We then iterated over a range of 64 period-amplitude pairs, injected a sinusoid with each pair of parameters into our light curve, and from this injected curve generated a Lomb-Scargle periodogram and bootstrapped detection limits. Finally, we measured the detection strength of our injected signal as the maximum peak of our periodogram divided by the periodogram's standard deviation, and plotted these strengths as a colormap. Amplitude is measured relative to the mean brightness of the base light curve. We overplot a linear interpolation of the $3\sigma$ detection limit, at each frequency bin. This limit has a maximum amplitude around a frequency of 0.01 Hz., and a small degree of bimodality.}
\end{center}
\end{figure}

\section{Acknowledgments}

The authors gratefully acknowledge the support of the
Caltech Optical Observatories and the Scialog workshop and grant. G.H. acknowledges the support of the National Science Foundation Career Award AST-1654815. K.J.S. is supported by NASA through the Astrophysics Theory Program (NNX17AG28G). B.D.M. is supported by the National Science Foundation (AST-1615084) and NASA (HST-AR-15041.001-A, GG008699). Laura Chomiuk is supported by a Cotrell Scholar Award of the Research Corporation and NSF-AST 1751874. The authors acknowledge with thanks the variable star observations from the AAVSO International Database contributed by observers worldwide and used in this research.

\newpage

\bibliographystyle{apj}

\bibliography{novabib}

\nocite{*}

\end{document}